\pdfoutput=1
\RequirePackage{ifpdf}
\ifpdf % We are running pdfTeX in pdf mode
\documentclass[pdftex]{sigma}
\else
\documentclass{sigma}
\fi

\usepackage{accents,multirow}

\numberwithin{equation}{section}

\newcommand{\wt}{\tilde}
\newcommand{\wh}{\hat}
\renewcommand{\th}[1]{\wh{\wt{#1}}}

\newcommand{\cP}{\mathcal{P}}
\newcommand{\cH}{\mathcal{H}}
\newcommand{\cB}{\mathcal{B}}

\newcommand{\mm}{\mathsf{m}}
\newcommand{\oT}{\mathsf{T}}

\newcommand{\sn}{\mathop{\mathrm{sn}}\nolimits}
\newcommand{\cn}{\mathop{\mathrm{cn}}\nolimits}
\newcommand{\dn}{\mathop{\mathrm{dn}}\nolimits}

\newcommand{\dt}{\circle*{2.5}}
\newcommand{\st}{\linethickness{1.0pt}}
\newcommand{\ct}{\linethickness{0.25pt}}
\newcommand{\nt}{\linethickness{0.1pt}}

\newcommand{\au}{\nt\vector(0,1){10}}
\newcommand{\ad}{\nt\vector(0,-1){10}}
\newcommand{\ar}{\nt\vector(1,0){10}}
\newcommand{\al}{\nt\vector(-1,0){10}}

\begin{document}

\allowdisplaybreaks

\renewcommand{\thefootnote}{$\star$}

\renewcommand{\PaperNumber}{073}

\FirstPageHeading

\ShortArticleName{Singularities of Type-Q ABS Equations}

\ArticleName{Singularities of Type-Q ABS Equations\footnote{This
paper is a contribution to the Proceedings of the Conference ``Symmetries and Integrability of Dif\/ference Equations (SIDE-9)'' (June 14--18, 2010, Varna, Bulgaria). The full collection is available at \href{http://www.emis.de/journals/SIGMA/SIDE-9.html}{http://www.emis.de/journals/SIGMA/SIDE-9.html}}}

\Author{James ATKINSON}

\AuthorNameForHeading{J.~Atkinson}

\Address{School of Mathematics and Statistics, The University of Sydney, NSW 2006, Australia}
\Email{\href{mailto:james.atkinson@sydney.edu.au}{james.atkinson@sydney.edu.au}}

\ArticleDates{Received February 14, 2011, in f\/inal form July 13, 2011;  Published online July 20, 2011}

\Abstract{The type-Q equations lie on the top level of the hierarchy introduced by Adler, Bobenko and Suris (ABS) in their classif\/ication of discrete counterparts of KdV-type integrable partial dif\/ferential equations.
We ask what singularities are possible in the solutions of these equations, and examine the relationship between the singularities and the principal integrability feature of multidimensional consistency.
These questions are considered in the global setting and therefore extend previous considerations of singularities which have been local.
What emerges are some simple geometric criteria that determine the allowed singularities, and the interesting discovery that generically the presence of singularities leads to a breakdown in the global consistency of such systems despite their local consistency pro\-perty.
This failure to be globally consistent is quantif\/ied by introducing a natural notion of monodromy for isolated singularities.}


\Classification{35Q58}

\renewcommand{\thefootnote}{\arabic{footnote}}
\setcounter{footnote}{0}

\section{Introduction}

The motivation for the present work comes from a desire to understand non-generic behaviour of the multidimensionally consistent systems listed in \cite{abs1} (cf.~\cite{nw,bs}) which can arise due to the presence of singularities.
Singularities in solutions of these systems have been considered previously.
A class of solutions which are singular {\it everywhere} have been known since~\cite{ahn}, their main interesting feature being a failure to signif\/icantly alter after application of the B\"acklund transformation.
Also recall that a local consideration of singularities was found in~\cite{abs2} to underlie the classif\/ication of consistent polynomials in terms of which the multidimensionally consistent systems are def\/ined, in particular providing an explanation of the spectral curve and the natural parameterisation of these systems in terms of points on that curve.
The objective here is dif\/ferent, namely to understand the full array of possible singularities in a {\it global} sense.

The local analysis of ABS in~\cite{abs2} showed that singularities in the solutions of these equations are naturally associated with the edges of the lattice (specif\/ically with the vanishing of edge biquadratics).
This gives rise to the question of how to determine globally the collections of edges that are admissible singularities.
There turn out to be a rich variety of possibilities, singularities can be isolated or they can form barriers partitioning the lattice, and their precise shape, or {\it configuration}, can vary a lot.
The f\/irst main outcome of this work is a set of geometric criteria which provide a basic test for the admissibility of a singularity conf\/iguration.

The general theory of the ABS equations often relies on being able to apply the B\"acklund transformation to a given solution, or in other words to extend the solution into a new lattice direction.
The second half of this paper (Section~\ref{ISM}) will describe the obstruction singularities can present to performing this operation.
Demanding a~solution be entirely free of singularities is a~rather strong requirement and it is natural to expect that something weaker would be suf\/f\/icient to guarantee the consistency of the B\"acklund equations, which is indeed the case.
The main problematic singularities are those which are isolated, their ef\/fect is quantif\/ied by introducing a notion of {\it monodromy}.
This can be used to resolve the global inconsistency of the B\"acklund equations, either by excluding from consideration those solutions whose singularities have non-trivial monodromy, or more interestingly through a~relaxation of the def\/inition of the domain of the transformed solution by allowing a modif\/ication of the lattice combinatorics.

The interaction between the B\"acklund transformation and singularities which form barriers is also of interest, in particular leading to a natural class of reductions of these systems.
The main issues there are somewhat dif\/ferent and will therefore be the subject of a separate publication.

\section{The class of equations and singularities in the solutions}\label{CLASS}

The focus here will be on the equations of type-Q def\/ined in~\cite{abs2}.
This is a restricted set of the discrete KdV-type equations, and this restriction requires some justif\/ication.
Specif\/ically, equations which are not of type-Q have the additional feature of {\it singular values}, special points in $\mathbb{C}\cup\{\infty\}$ (and usually canonical forms dictate these are $0$ or $\infty$) whose presence in a solution indicates a singularity.
The type-Q equations {\it have no such special points}, however it will be seen that they do have a rich theory of singularities, many aspects of which translate to the equations which are not of type-Q.
The presence of singular values mean the non-Q (or type-H) equations deserve a separate treatment, the precise singularity structure being essentially dif\/ferent.
\subsection{Basic def\/initions}
The equations to be considered take the form
\begin{gather} \cP(u,\wt{u},\wh{u},\th{u})=0, \label{ge} \end{gather}
where $\cP$ is one of the polynomials listed in Table~\ref{Qpolys}.
$u=u(n,m)$, $\wt{u}=u(n+1,m)$, $\wh{u}=u(n,m+1)$ and $\th{u}=u(n+1,m+1)$ are values of the dependent variable $u$ as a function of independent variables $n,m\in\mathbb{Z}$, therefore equation (\ref{ge}) connects values of $u$ on the vertices of each quadrilateral of the $\mathbb{Z}^2$ lattice.
Our intention is to describe singularities for all equations of the form~(\ref{ge}) def\/ined by type-Q polynomials with Kleinian symmetry, but according to~\cite{abs2} we can restrict attention to the canonical forms given in Table~\ref{Qpolys} without losing generality.
\begin{table}[t]\centering
\caption{Canonical forms for the type-Q polynomials. $\alpha$, $\beta$ are complex parameters, $\delta\in\{0,1\}$ and $k\in\mathbb{C}\setminus\{-1,0,1\}$ is the modulus of the Jacobi $\sn$ function appearing in Q4.
The given function $f$ serves to parameterise a curve $(f')^2=r(f)$ def\/ined in terms of a polynomial~$r$, which loosely speaking serves to characterise the equation itself \cite{adl,abs2}.}
\label{Qpolys}
\vspace{1.5mm}

\begin{tabular}{cll}
\hline
& \tsep{2pt} \tsep{1pt} $\cP(u,\wt{u},\wh{u},\th{u})$ & $f(\zeta)$\\
\hline
 \multirow{2}{*}{Q4\hphantom{$^\delta$}}&
$\sn(\alpha)(u\wt{u}+\wh{u}\wh{\wt{u}})-\sn(\beta)(u\wh{u}+\wt{u}\wh{\wt{u}})$ & \multirow{2}{*}{$\sqrt{k}\sn(\zeta)$}\tsep{2pt}\\
& $\quad-\sn(\alpha-\beta)[\wt{u}\wh{u}+u\wh{\wt{u}}-k\sn(\alpha)\sn(\beta)(1+u\wt{u}\wh{u}\wh{\wt{u}})]$ \\
\multirow{2}{*}{Q3$^\delta$}&
$\sinh(\alpha)(u\wt{u}+\wh{u}\th{u})-\sinh(\beta)(u\wh{u}+\wt{u}\th{u})$ & \multirow{2}{*}{$\frac{1}{2}[\exp(\zeta)+\delta^2\exp(-\zeta)]$}\\
&$\quad-\sinh(\alpha-\beta)(\wt{u}\wh{u}+u\th{u}+\delta^2\sinh(\alpha)\sinh(\beta))$\\
\multirow{2}{*}{Q2\hphantom{$^\delta$}}&
$\alpha(u-\wh{u})(\wt{u}-\th{u})-\beta(u-\wt{u})(\wh{u}-\th{u})$ & \multirow{2}{*}{$\zeta^2$}\\
&$\quad+\alpha\beta(\alpha-\beta)(u+\wt{u}+\wh{u}+\th{u}-\alpha^2+\alpha\beta-\beta^2)$\\
Q1$^\delta$&
$\alpha(u-\wh{u})(\wt{u}-\th{u})-\beta(u-\wt{u})(\wh{u}-\th{u})+\delta^2 \alpha\beta(\alpha-\beta)$ & $\delta\zeta$\\
\hline
\end{tabular}
\end{table}

Consider numbering the four vertices of a quadrilateral in the order the corresponding variables appear in the argument of~$\cP$ in~(\ref{ge}).
This allows the def\/inition:
\begin{definition}\label{singdef}\qquad
\begin{enumerate}\itemsep=0pt
\item[$(i)$] A solution of (\ref{ge}) is said to be singular with respect to vertex $i$ of some quadrilateral if
\begin{gather*}
[\partial_i\cP](u,\wt{u},\wh{u},\th{u})=0
%\label{sc}
\end{gather*}
also holds on that quadrilateral.
\item[$(ii)$] The set of all quadrilaterals on which the solution is singular is called the singular region.
\end{enumerate}
Here $\partial_i\cP$ denotes the partial derivative of $\cP$ with respect to its $i^{\rm th}$ argument, so actually has no dependence on the $i^{\rm th}$ argument because $\cP$ is a polynomial of degree one in each argument.
This def\/inition means a solution is singular with respect to vertex $i$ on some quadrilateral when it satisf\/ies (\ref{ge}) independently of its value on vertex~$i$.
\end{definition}

As an explicit example consider the solution mentioned in the introduction which is singular everywhere.
This solution is of the form
\begin{gather}
u=f(\zeta)\label{ssol}
\end{gather}
with $f$ written in Table~\ref{Qpolys} and $\zeta$ a function def\/ined on $\mathbb{Z}^2$ by the system
\begin{gather}
\wt{\zeta}=\zeta+\epsilon_1\alpha, \qquad \wh{\zeta}=\zeta+\epsilon_2\beta,\label{zetasys}
\end{gather}
where
\begin{gather*}
\epsilon_1,\epsilon_2: \ \mathbb{Z}^2\longrightarrow\{-1,+1\}, \qquad \wt{\epsilon}_2=\epsilon_2, \qquad \wh{\epsilon}_1=\epsilon_1.
\end{gather*}
The functions $\epsilon_1$ and $\epsilon_2$ are constant in one lattice direction so that the system (\ref{zetasys}) is compatible, and they attach an arbitrary $+1$ or $-1$ to the edges of the lattice in the other lattice direction.
It can be directly verif\/ied that the resulting function $u$ def\/ined in (\ref{ssol}) is a solution of the equation which is singular with respect to two diagonally opposite vertices on every quadrilateral (the relative signs of $\epsilon_1$ and $\epsilon_2$ determine which diagonal pair).

\subsection{Singular edges}

It can be seen from Def\/inition \ref{singdef} that to locate a singularity requires the specif\/ication of a~quadrilateral, as well as a particular vertex of that quadrilateral.
However the description of singularities can be simplif\/ied by considering the polynomials
\begin{gather}
\cH_{ij} = (\partial_i\cP)(\partial_j\cP)-(\partial_i\partial_j\cP)\cP, \qquad i,j\in\{1,2,3,4\}, \quad i\neq j.\label{biquad}
\end{gather}
Clearly $\cH_{ij}$ is invariant under interchange of the indices $i$ and $j$, also it is not dif\/f\/icult to verify that $\cH_{ij}$, considered as a polynomial in four variables, is degree zero in arguments $i$ and $j$ (has no dependence on argument $i$ or $j$).
It follows from the invariance of $\cP$ under Kleinian permutations of its arguments (this invariance is easily verif\/ied by glancing at Table \ref{Qpolys}) that there are really only three polynomials def\/ined by (\ref{biquad}), specif\/ically
\begin{gather}
h_{12}(x,y):=\cH_{12}(\cdot,\cdot,x,y)=\cH_{34}(x,y,\cdot,\cdot),\nonumber\\
h_{13}(x,y):=\cH_{13}(\cdot,x,\cdot,y)=\cH_{24}(x,\cdot,y,\cdot),\label{bred}\\
h_{14}(x,y):=\cH_{14}(\cdot,x,y,\cdot)=\cH_{23}(x,\cdot,\cdot,y).\nonumber
\end{gather}
These are the symmetric biquadratic polynomials which play a basic role in the theory of~(\ref{ge}) developed in \cite{adl,abs1,abs2}.
Explicit forms of the biquadratics are given in Table~\ref{Qhs}.
The most important aspect here is that, excluding $Q1^{0}$, these biquadratics are the addition formulae for the functions $f$ appearing in Table~\ref{Qpolys}, specif\/ically
\begin{gather}
 h_{12}(f(\zeta),f(\wt{\zeta}))=0 \quad \Leftrightarrow \quad f(\wt{\zeta})=f(\zeta\pm\alpha),\nonumber\\
 h_{13}(f(\zeta),f(\wh{\zeta}))=0 \quad \Leftrightarrow \quad f(\wh{\zeta})=f(\zeta\pm\beta),\label{bqd}\\
 h_{14}(f(\zeta),f(\th{\zeta}))=0 \quad \Leftrightarrow \quad f(\th{\zeta})=f(\zeta\pm\alpha\mp\beta).\nonumber
\end{gather}
\begin{table}[t]
\centering
\caption{Biquadratics $h_{12}$ obtained from the polynomials listed in Table~\ref{Qpolys}. The other biquadratics can be obtained from these by the equations $h_{13}=h_{12}\vert_{\alpha\leftrightarrow\beta}$ and $h_{14}=h_{12}\vert_{\alpha\rightarrow\beta-\alpha}$.}\label{Qhs}
\vspace{1.5mm}

\begin{tabular}{cl}
\hline
& $h_{12}(x,y)$ \\
\hline
Q4\hphantom{$^\delta$}& $\sn(\beta)\sn(\alpha-\beta)[x^2+y^2-(1+x^2y^2)\sn^2(\alpha)-2xy\cn(\alpha)\dn(\alpha)]$\tsep{2pt}\\
Q3$^\delta$& $\sinh(\beta)\sinh(\alpha-\beta)[(y-x e^{\alpha})(y-xe^{-\alpha})+\delta^2\sinh^2(\alpha)]$\\
Q2\hphantom{$^\delta$}& $\beta(\alpha-\beta)[y^2+x^2-2xy-2\alpha^2(x+y)+\alpha^4]$\\
Q1$^\delta$& $\beta(\alpha-\beta)(y-x-\delta \alpha)(y-x+\delta\alpha)$\\
\hline
\end{tabular}
\end{table}

From Def\/inition \ref{singdef}$(i)$ and the def\/inition of $\cH_{ij}$ in (\ref{biquad}) it can be seen that
\begin{lemma}[ABS {\cite{abs2}}]\label{absl3}
A solution of \eqref{ge} satisfies $\cH_{ij}(u,\wt{u},\wh{u},\th{u})=0$ on some quadrilateral if and only if it is singular with respect to vertex $i$ or $j$ of that quadrilateral.
\end{lemma}

 This connection between singularities and the vanishing of the biquadratics is key in what follows.

\begin{figure}[t]
\centering
\begin{picture}(100,100)(0,0)
\multiput(21,12)(60,0){2}{{\st\line(0,1){60}}}
\multiput(21,12)(0,60){2}{{\st\line(1,0){60}}}
\multiput(0,39)(84,0){2}{{$h_{13}$}}
\multiput(43,0)(0,78){2}{{$h_{12}$}}
\put(21,12){\circle*{3}}
\put(81,12){\circle*{3}}
\put(21,72){\circle*{3}}
\put(81,72){\circle*{3}}
\put(12,4){$u$}
\put(84,4){$\wt{u}$}
\put(12,72){$\wh{u}$}
\put(84,72){$\th{u}$}
\end{picture}
\caption{Values of the dependent variable associated to the vertices of a quadrilateral and biquadratics def\/ined in~(\ref{biquad}),~(\ref{bred}) associated to the edges.}\label{quad}
\end{figure}
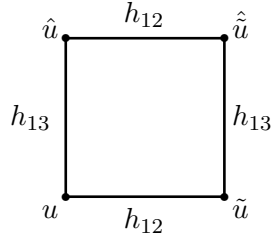

Biquadratics $h_{12}$ and $h_{13}$ are naturally associated with the edges of the quadrilateral as shown in Fig.~\ref{quad}\footnote{The remaining biquadratic $h_{14}$ is naturally associated with the diagonals.}.
And because the biquadratics on opposite edges of the quadrilateral coincide, they can be associated in a consistent way to all edges of the lattice.
The problem of describing singularities reduces to listing the edges of the lattice on which the biquadratics vanish, thus it is natural to make the following def\/inition:
\begin{definition}\label{se}
A solution of (\ref{ge}) is said to be singular on some edge of the lattice if the biquadratic on that edge (cf. Fig.~\ref{quad}) vanishes.
\end{definition}

It then follows from Def\/inition \ref{singdef}$(ii)$ and Lemma~\ref{absl3} that
\begin{lemma}\label{er}
A quadrilateral is part of the singular region $($cf.\ Definition {\rm \ref{singdef}$(ii)$)} if and only if it has a singular edge.
\end{lemma}

But not every collection of edges is an admissible singularity conf\/iguration.

\subsection{Singularity conf\/iguration conditions}\label{SC}

Given a collection of edges we ask if they constitute an admissible singularity conf\/iguration.
Constraints on admissible conf\/igurations can be broken down into the three conditions which follow.
The f\/irst condition should be modif\/ied and the second removed altogether for equation~$Q1^0$, which is a special case whose solution is {\it constant} on vertices connected by singular edges.

\begin{figure}[t]
\centering
\begin{picture}(240,120)(0,0)
% Part (a)
\put(0,91){${(a)}$}
\multiput(38,78)(80,0){3}{\circle*{2}}
\multiput(68,78)(80,0){3}{\circle*{2}}
\multiput(68,108)(80,0){3}{\circle*{2}}
\multiput(38,108)(80,0){3}{\circle*{2}}
\multiput(38,78)(80,0){3}{\ct\line(1,0){30}}
\multiput(38,78)(80,0){3}{\ct\line(0,1){30}}
\multiput(148,78)(80,0){2}{\ct\line(0,1){30}}
\put(198,108){\ct\line(1,0){30}}
% Part (b), as (a) but with arrows
\put(0,21){${(b)}$}
\multiput(38,8)(80,0){3}{\circle*{2}}
\multiput(68,8)(80,0){3}{\circle*{2}}
\multiput(68,38)(80,0){3}{\circle*{2}}
\multiput(38,38)(80,0){3}{\circle*{2}}
\multiput(38,8)(80,0){3}{\ct\line(1,0){30}}
\multiput(38,8)(80,0){3}{\ct\vector(1,0){18}}
\multiput(38,8)(80,0){3}{\ct\line(0,1){30}}
\multiput(38,8)(80,0){3}{\ct\vector(0,1){18}}
\multiput(148,8)(80,0){2}{\ct\line(0,1){30}}
\put(148,38){\ct\vector(0,-1){18}}
\put(198,38){\ct\line(1,0){30}}
\put(228,8){\ct\vector(0,1){18}}
\put(198,38){\ct\vector(1,0){18}}
\end{picture}
\caption{Admissible singularity conf\/igurations on a single quadrilateral in part $(a)$, and attached arrows in part $(b)$. The solid lines indicate singular edges. The conf\/igurations are given up to the symmetries of the square and also up to an overall reversal of arrows in part~$(b)$. In the exceptional case of equation~$Q1^0$ the second conf\/iguration involving three singular edges is inadmissible.}
\label{ss}
\end{figure}
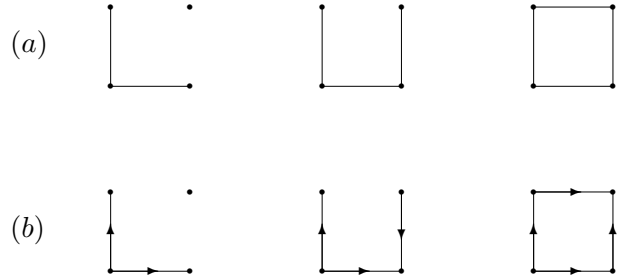

The f\/irst condition concerns each quadrilateral within the singular region.
It is clear from Lemma~\ref{absl3} that if a solution of (\ref{ge}) is singular on some edge of a quadrilateral, then it must also be singular on one of the two edges adjacent to the singular edge.
This means that each quadrilateral within the singular region should be one of those shown in Fig.~\ref{ss}$(a)$.

The second condition concerns each connected singular region.
From Def\/inition \ref{se} and relations (\ref{bqd}) the solution on a set of vertices connected by singular edges is of the form $u=f(\zeta)$ where $\wt{\zeta}=\zeta+\epsilon_1\alpha$, $\wh{\zeta}=\zeta+\epsilon_2\beta$ and $\epsilon_1,\epsilon_2:\mathbb{Z}^2\rightarrow\{-1,+1\}$.
The functions $\epsilon_1$ and $\epsilon_2$ attach a~$+1$ or~$-1$ to each singular edge, it is convenient to indicate this function geometrically as an arrow in the positive or negative lattice direction respectively.
An analysis of~(\ref{bqd}) and Lemma~\ref{absl3} reveals that on each quadrilateral there is a constraint on these arrows, specif\/ically they must be as shown in Fig.~\ref{ss}$(b)$.
Furthermore, it is necessary that around any closed path of singular edges the arrows sum to the null vector.
This further condition is necessary for the compatibility of the aforementioned system for~$\zeta$ and leads to explicit construction of the solution on the connected singular region.

Finally, supposing that the solution can be constructed on the singular region, we ask if a~solution can then be constructed on the {\it non-singular} region.
The only obstruction to this could be that the solution on this region is overdetermined.
This can be checked at f\/irst by comparing the number of free vertices (i.e.\ vertices not attached to a singular edge) with the number of quadrilaterals (i.e.\ the number of constraints).
If the solution on the non-singular region appears overdetermined by this test then further consideration is necessary because it might happen that some symmetry of the problem resolves the apparent inconsistency.

To summarise these conditions we formulate the following theorem.
\begin{theorem}\label{conds}
Propose a solution of \eqref{ge} with some collection of singular edges $($cf.\ Definition~{\rm \ref{se})} which determine some singular region $($cf.\ Lemma~{\rm  \ref{er})}.
Sufficient conditions for existence of this solution are as follows:
\begin{itemize}\itemsep=0pt
\item[$1$.]
Each quadrilateral within the singular region should be one of those in Fig.~{\rm \ref{ss}$(a)$}.
\item[$2$.]
It should be possible to attach arrows to all singular edges in such a way that $(i)$ each quadrilateral within the singular region is one of those shown in Fig.~{\rm \ref{ss}$(b)$} and~$(ii)$ around any closed path of singular edges the arrows sum to the null vector\footnote{Condition~2 does not apply to equation $Q1^0$.}.
\item[$3$.]
The solution on the non-singular region should not be overdetermined.
\end{itemize}
\end{theorem}

Singularity conf\/igurations which violate condition 2 are possible but are non-generic.
Such conf\/igurations would require additional constraints connecting some or all of $\alpha$, $\beta$, the periodicity of $f$ (which are intrinsic to the equation) and the integration constant $\zeta_0$ on each connected singular region.

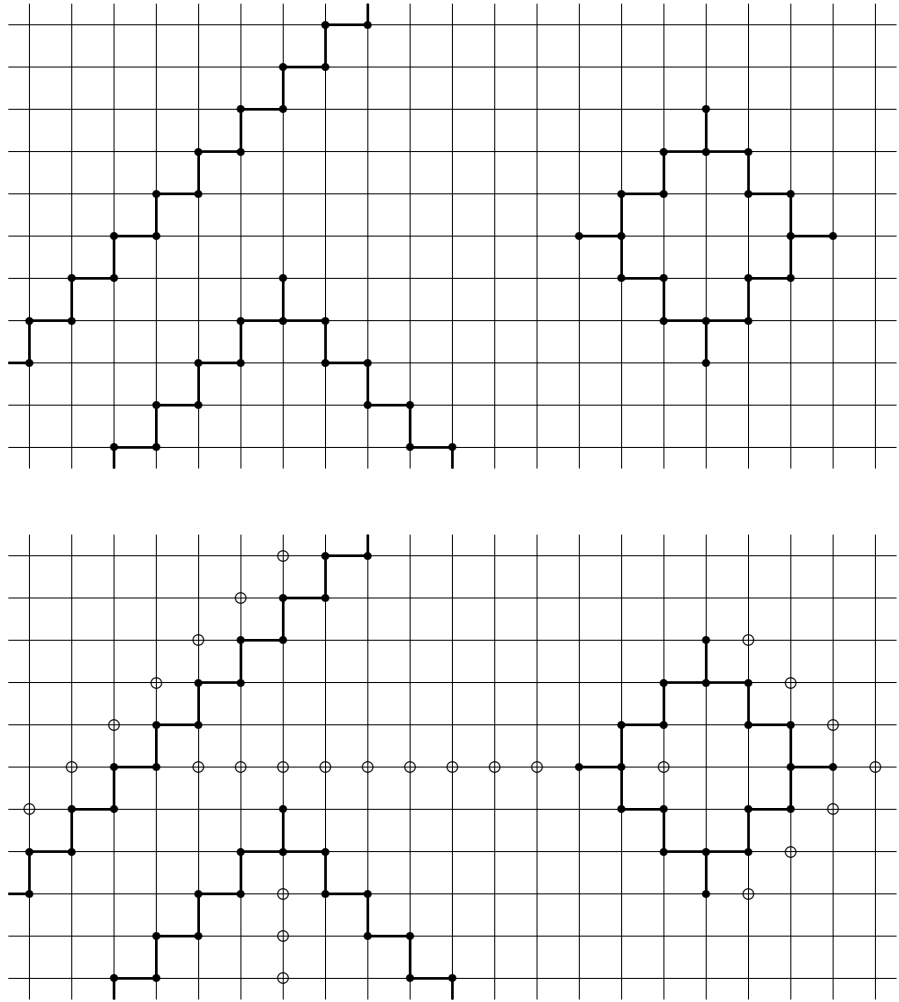
\begin{figure}[th!]
\centering
\begin{picture}(340,180)(0,0)
% 10x20 grid, 16x16 squares
\multiput(10,2)(16,0){21}{{\nt\line(0,1){176}}}
\multiput(2,10)(0,16){11}{{\nt\line(1,0){336}}}
% infringement-of-copyright-staircase
\multiput(10,42)(16,16){9}{\dt}
\multiput(10,58)(16,16){8}{\dt}
\multiput(10,58)(16,16){8}{{\st\line(1,0){16}}}
\multiput(10,42)(16,16){8}{{\st\line(0,1){16}}}
\put(2,42){{\st\line(1,0){8}}}
\put(138,170){{\st\line(0,1){8}}}
% triangle
\put(42,2){{\st\line(0,1){8}}}
\multiput(42,10)(16,16){5}{\dt}
\multiput(58,10)(16,16){4}{\dt}
\multiput(42,10)(16,16){4}{{\st\line(1,0){16}}}
\multiput(58,10)(16,16){4}{{\st\line(0,1){16}}}
\put(170,2){{\st\line(0,1){8}}}
\multiput(170,10)(-16,16){4}{\dt}
\multiput(154,10)(-16,16){3}{\dt}
\multiput(170,10)(-16,16){4}{{\st\line(-1,0){16}}}
\multiput(154,10)(-16,16){3}{{\st\line(0,1){16}}}
% transparent diamond
\multiput(266, 42)( 16, 16){3}{\dt}
\multiput(314, 90)(-16, 16){3}{\dt}
\multiput(266,138)(-16,-16){3}{\dt}
\multiput(218, 90)( 16,-16){3}{\dt}
\multiput(266, 58)( 16, 16){2}{\dt}
\multiput(298, 90)(-16, 16){2}{\dt}
\multiput(266,122)(-16,-16){2}{\dt}
\multiput(234, 90)( 16,-16){2}{\dt}
\multiput(266, 42)( 16, 16){3}{{\st\line(0,1){16}}}
\multiput(314, 90)(-16, 16){3}{{\st\line(-1,0){16}}}
\multiput(266,138)(-16,-16){3}{{\st\line(0,-1){16}}}
\multiput(218, 90)( 16,-16){3}{{\st\line(1,0){16}}}
\multiput(266, 58)( 16, 16){2}{{\st\line(1,0){16}}}
\multiput(298, 90)(-16, 16){2}{{\st\line(0,1){16}}}
\multiput(266,122)(-16,-16){2}{{\st\line(-1,0){16}}}
\multiput(234, 90)( 16,-16){2}{{\st\line(0,-1){16}}}
\end{picture}
\begin{picture}(340,200)(0,0)
% 10x20 grid, 16x16 squares
\multiput(10,2)(16,0){21}{{\nt\line(0,1){176}}}
\multiput(2,10)(0,16){11}{{\nt\line(1,0){336}}}
% infringement-of-copyright-staircase
\multiput(10,42)(16,16){9}{\dt}
\multiput(10,58)(16,16){8}{\dt}
\multiput(10,58)(16,16){8}{{\st\line(1,0){16}}}
\multiput(10,42)(16,16){8}{{\st\line(0,1){16}}}
\put(2,42){{\st\line(1,0){8}}}
\put(138,170){{\st\line(0,1){8}}}
% triangle
\put(42,2){{\st\line(0,1){8}}}
\multiput(42,10)(16,16){5}{\dt}
\multiput(58,10)(16,16){4}{\dt}
\multiput(42,10)(16,16){4}{{\st\line(1,0){16}}}
\multiput(58,10)(16,16){4}{{\st\line(0,1){16}}}
\put(170,2){{\st\line(0,1){8}}}
\multiput(170,10)(-16,16){4}{\dt}
\multiput(154,10)(-16,16){3}{\dt}
\multiput(170,10)(-16,16){4}{{\st\line(-1,0){16}}}
\multiput(154,10)(-16,16){3}{{\st\line(0,1){16}}}
% transparent diamond
\multiput(266, 42)( 16, 16){3}{\dt}
\multiput(314, 90)(-16, 16){3}{\dt}
\multiput(266,138)(-16,-16){3}{\dt}
\multiput(218, 90)( 16,-16){3}{\dt}
\multiput(266, 58)( 16, 16){2}{\dt}
\multiput(298, 90)(-16, 16){2}{\dt}
\multiput(266,122)(-16,-16){2}{\dt}
\multiput(234, 90)( 16,-16){2}{\dt}
\multiput(266, 42)( 16, 16){3}{{\st\line(0,1){16}}}
\multiput(314, 90)(-16, 16){3}{{\st\line(-1,0){16}}}
\multiput(266,138)(-16,-16){3}{{\st\line(0,-1){16}}}
\multiput(218, 90)( 16,-16){3}{{\st\line(1,0){16}}}
\multiput(266, 58)( 16, 16){2}{{\st\line(1,0){16}}}
\multiput(298, 90)(-16, 16){2}{{\st\line(0,1){16}}}
\multiput(266,122)(-16,-16){2}{{\st\line(-1,0){16}}}
\multiput(234, 90)( 16,-16){2}{{\st\line(0,-1){16}}}
% initial data
\multiput(10,74)(16,16){7}{\circle{4}}
\multiput(74,90)(16,0){9}{\circle{4}}
\multiput(106,10)(0,16){3}{\circle{4}}
\multiput(330,90)(-16,16){4}{\circle{4}}
\multiput(314,74)(-16,-16){3}{\circle{4}}
\put(250,90){\circle{4}}
\end{picture}
\caption{Example of an admissible conf\/iguration of singular edges in the solution of a type-Q lattice equation in two dimensions. The f\/irst diagram shows the singular edges which clearly satisfy condition~1 of Theorem~\ref{conds}. It is a simple exercise to attach arrows to the singular edges as described in condition 2. The second diagram is the same as the f\/irst but also shows an example set of vertices on which arbitrary non-singular initial data can be specif\/ied from which a unique solution is determined, thus demonstrating the conf\/iguration passes also condition 3. Within the region completely enclosed by singularities there are f\/ive variables and four equations, so essentially one degree of freedom remains which is indicated by the single vertex on which initial data can be specif\/ied, actually there are still two solutions possible in this region so it is not quite determined uniquely here.}
\label{admissible}
\end{figure}

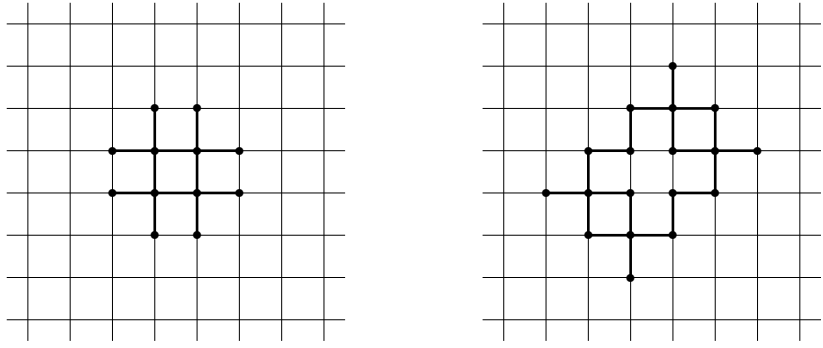
\begin{figure}[t]
\centering
\begin{picture}(320,140)(0,0)
% configuration fails on condition 2
\multiput(10,2)(16,0){8}{{\nt\line(0,1){128}}}
\multiput(2,10)(0,16){8}{{\nt\line(1,0){128}}}
\multiput(58,42)(0,16){4}{\multiput(0,0)(16,0){2}{\dt}}
\multiput(42,58)(48,0){2}{\multiput(0,0)(0,16){2}{\dt}}
\multiput(58,42)(16,0){2}{\st\line(0,1){48}}
\multiput(42,58)(0,16){2}{\st\line(1,0){48}}
% configuration fails on condition 3
\multiput(190,2)(16,0){8}{{\nt\line(0,1){128}}}
\multiput(182,10)(0,16){8}{{\nt\line(1,0){128}}}
\multiput(238,26)(16,16){4}{\dt}
\multiput(238,42)(16,16){3}{\dt}
\multiput(222,42)(16,16){4}{\dt}
\multiput(222,58)(16,16){3}{\dt}
\multiput(206,58)(16,16){4}{\dt}
\multiput(222,58)(16,16){3}{\st\line(0,1){16}}
\multiput(222,58)(16,16){3}{\st\line(-1,0){16}}
\multiput(238,42)(16,16){3}{\st\line(0,-1){16}}
\multiput(238,42)(16,16){3}{\st\line(1,0){16}}
\multiput(222,58)(0,-16){2}{\st\line(1,0){16}}
\multiput(254,90)(0,-16){2}{\st\line(1,0){16}}
\multiput(222,58)(16,0){2}{\st\line(0,-1){16}}
\multiput(254,90)(16,0){2}{\st\line(0,-1){16}}
\end{picture}
\caption{Collections of edges which fail the singularity conf\/iguration conditions of Theorem~\ref{conds}. It is easily checked that the f\/irst conf\/iguration passes condition~1 and~2$(i)$ but fails on condition 2$(ii)$, whilst the second conf\/iguration passes conditions~1 and~2 but a calculation shows that generically it fails on condition~3.}
\label{not-admissible}
\end{figure}

The conditions given in Theorem \ref{conds} make it possible to geometrically determine admissible singularity conf\/igurations, an example is illustrated in Fig.~\ref{admissible}.
Examples of (generically) inadmissible conf\/igurations which fail on conditions 2 and 3 respectively of Theorem~\ref{conds} are shown in Fig.~\ref{not-admissible}.
It is necessary to account for the possibility of failure on condition 3, however to conceive of an example took some time.
In general a conf\/iguration which passes conditions 1 and 2 but which fails on condition 3 is rare, the example given on the right in Fig.~\ref{not-admissible} should be considered pathological.

There is an obvious algorithm to test these conditions, the algorithm would use a f\/inite number of operations provided the singular region were f\/inite.
This is just a consequence of the equations being discrete; this determinism is not connected to the integrability.
However, the Kleinian symmetry of the def\/ining polynomials was essentially what led to the consistent association of singularities to edges of the lattice.
This association is perhaps the most critical feature of the singularities because it simplif\/ies many-fold their description.
But the Kleinian symmetry is synonymous with the integrability for these systems, so the basic simplif\/ication of the problem af\/forded by association of singularities to edges is tied intimately to the integrability of the equations.

\subsection{Extension to quad-graphs and the multidimensional lattice}\label{mdc}

This paper so-far has been focused on equations def\/ined on the regular two-dimensional quadrilateral lattice, but the same equations can be considered in a more general setting.
The restriction has been made in order to introduce the underlying ideas in as simple a setting as possible.
In practice it is important to also consider the following two well-known generalisations.

First, the equations can be considered on the more general edge-labelled {\it quad-graph} as considered in~\cite{bs}.
It is natural to impose a mild technical assumption on the graph, which was established in~\cite{av}, that characteristics do not self-intersect.
This is because the system is basically unchanged by removing quadrilaterals on which the self-intersections occur.

Second, the equations can be considered on the regular {\it multidimensional} lattice.
The extension of the underlying equation (\ref{ge}) to a system in $N$ dimensions is
\begin{gather}
\cP_{ij}(u,\oT_iu,\oT_ju,\oT_i\oT_ju)=0, \qquad i,j\in\{1,\ldots, N\}, \quad i\neq j,\label{mde}
\end{gather}
where $\cP_{ij}=\cP|_{(\alpha,\beta)\rightarrow(\alpha_i,\alpha_j)}$, $u$ takes values on the vertices of the $N$-dimensional regular lattice, $\oT_1,\ldots,\oT_N$ are shift operators in the $N$ dimensions, and $\alpha_1,\ldots,\alpha_N$ are a set of parameters\footnote{Actually, the generalisation to multidimensions {\it contains} the generalisation to quad-graphs because every quad-graph with non self-intersecting characteristics can be embedded in the {\it regular} multidimensional lattice~\cite{dsss}. Such embedding was used to establish the natural Cauchy-problem for the multidimensionally consistent equations on the quad-graph in~\cite{av}.}.

In the previous sections only the examples given relied on the setting of the regular two-dimensional lattice, the def\/initions, lemmas and singularity conf\/iguration conditions have been stated in such a way that they are {\it unchanged} in the more general settings of quad-graph or regular multidimensional lattice.
The most important fact that makes this possible is that the basic association of singularities to {\it edges} of the lattice continues to be valid.
This relies on the result established in~\cite{abs1,abs2} that the biquadratics on edges shared by any pair of quadrilaterals coincide up to a constant factor (this is not always true for consistent systems involving type-H polynomials).
In conclusion, the singularity conf\/iguration conditions remain valid in both of these more general settings.

\section{Isolated singularities and monodromy}\label{ISM}

The consideration of singularities in multidimensions (cf.\ Section~\ref{mdc}) comes very close to fully describing the interaction between singularities and the B\"acklund transformation because applying the transformation means simply extending the solution one iterate into a new lattice direction.
But there is one aspect of this interaction which that point of view overlooks.
To describe this overlooked feature the question will be considered afresh, starting on the regular two-dimensional lattice (again, nothing important will be lost in this setting).

\subsection{The B\"acklund transformation}\label{TBT}
The B\"acklund transformation provides a reason why it is natural to associate singularities to the edges of the lattice.
The B\"acklund equations of~(\ref{ge}) take the form
\begin{gather}
\cB_1(u,\wt{u},v,\wt{v})=0,\qquad \cB_2(u,\wh{u},v,\wh{v})=0,\label{bt}
\end{gather}
where $\cB_1=\cP|_{(\alpha,\beta)\rightarrow(\alpha,\gamma)}$ and $\cB_2=\cP|_{(\alpha,\beta)\rightarrow(\beta,\gamma)}$ ($\cP$ was the polynomial appearing in the equation~(\ref{ge}) itself), they determine a new solution $v$ of~(\ref{ge}) from an old solution $u$, the free pa\-ra\-me\-ter~$\gamma$ is the B\"acklund parameter.
Because $\cB_1$ and $\cB_2$ are polynomials of degree one the system~(\ref{bt}) def\/ines M\"obius transformations $v\mapsto\wt{v}$ and $v\mapsto\wh{v}$ associated with the (oriented) edges of the lattice.
It happens that {\it $u$ is singular on some edge exactly when the M\"obius transformation is not defined on that edge}.
This occurs when (\ref{bt}) are reducible as polynomials in~$(v,\wt{v})$ and~$(v,\wh{v})$ respectively.

\subsection{Monodromy}\label{MON}

The global compatibility of (\ref{bt}) as a system for $v$ means that the composition of M\"obius transformations along some path connecting two vertices of the lattice should be independent of the path chosen.
Now, the path-independence of composition around a {\it single quadrilateral} is equivalent to consistency of the polynomials on the cube.
Path-independence of the composition in a global sense then follows when any path between two vertices can be deformed into any other path by changing it gradually {\it on one quadrilateral at a time}.
This argument for global compatibility breaks down in the presence of isolated singularities, i.e.\ localised collections of edges on which the M\"obius transformations are not def\/ined, because if the two paths enclose a~singularity they cannot be deformed one into the other in this way.

The possibility of an isolated singularity to result in global inconsistency of the B\"acklund system (\ref{bt}) can be quantif\/ied by introducing the {\it monodromy} of the singularity.
Consider composing transformations along a closed path of non-singular edges enclosing the singularity.
It is easily shown that deforming the path (without moving the point of origin) on one non-singular quadrilateral at a time leaves the composed transformation unchanged.
Whilst moving the point of origin or reversing the orientation of the path leads to conjugation of the composed transformation\footnote{The path-reversal of course leads to inversion of the transformation, but every M\"obius transformation is conjugate to its inverse.}.
Up to conjugacy this composed transformation is the natural monodromy of the singularity:
\begin{definition}\label{mon}\qquad
\begin{enumerate}\itemsep=0pt
\item[$(i)$] An isolated singularity in a solution of (\ref{ge}) is a collection of singular edges for which there exists a f\/inite-length enclosing path of non-singular edges.
\item[$(ii)$] Let $\mm_1,\ldots,\mm_N$ denote the M\"obius transformations associated by the B\"acklund equations~(\ref{bt}) to edges along a non-singular oriented path enclosing an isolated singularity such that the indices increase with distance along the path.
Then the monodromy of the singularity is def\/ined as the conjugacy class of the M\"obius transformation $\mm=\mm_{N}\cdot\mm_{N-1}\cdots \mm_1$.
\item[$(iii)$] If this transformation is the identity we say the monodromy is trivial.
\end{enumerate}
\end{definition}

 The monodromy is independent of the shape and orientation of the path.
If it is trivial the singularity does not lead to global inconsistency of the B\"acklund system.
It can be seen from Def\/inition~\ref{mon} and the explicit forms of the polynomials in Table~\ref{Qpolys} that the trivial-monodromy condition is polynomial in $f(\gamma)$ and $f'(\gamma)$, where~$\gamma$ is the B\"acklund parameter.

We remark that generically the potential for global inconsistency is always present for B\"ack\-lund transformations def\/ined by the polynomials consistent on a cube.
It happens whenever the B\"acklund equations fail to def\/ine the M\"obius transformations on some collection of edges, and therefore the monodromy can be def\/ined in the same way for any similar B\"acklund transformations, including those involving type-H polynomials.
There is, in principle, the possibility that singularities of the B\"acklund equations are distinct from singularities of the equation, this would indicate a kind of incompatibility between the singularity structure and the integrability, and it would be interesting to investigate to see if such systems exist (for instance amongst those listed in~\cite{boll}) and to f\/ind out how the incompatibility is resolved.

\subsection{Calculation of monodromy}

To calculate the monodromy it is useful to exploit that there is a single complex parameter which distinguishes the conjugacy classes in the M\"obius group.
Consider the non-identity M\"obius transformation $\mm=x\mapsto (ax+b)/(cx+d)$, where $a,b,c,d\in\mathbb{C}$, $ad\neq bc$.
Then the relationship between $\mm$ and the complex parameter $\rho$
\begin{gather}
\frac{(\rho+1)^2}{\rho} = \frac{(x-\mm^2(x))(\mm(x)-\mm^3(x))}{(x-\mm(x))(\mm^2(x)-\mm^3(x))} = \frac{(a+d)^2}{ad-bc},\label{rhodef}
\end{gather}
where the second expression is independent of the arbitrary variable $x$, has the following properties.
First, if $N\in\mathbb{Z}$ then $\mm\rightarrow\mm^N \Rightarrow \rho\rightarrow \rho^N$.
Second, if $\mm_*$ and $\rho_*$ are some other M\"obius transformation and parameter related by~(\ref{rhodef}), then $\{\rho_*,1/\rho_*\}=\{\rho,1/\rho\}$ if and only if~$\mm_*$ is conjugate to~$\mm$.
These properties can be verif\/ied directly, or alternatively they can be deduced from matrix properties by observing that $\rho$ is the ratio of the eigenvalues of any matrix in $PGL(2,\mathbb{C})$ corresponding to the M\"obius transformation $\mm$ by the natural homomorphism.
The expression~(\ref{rhodef}) determines $\rho$ up to the transformation $\rho\rightarrow 1/\rho$, which we identify with reversing the orientation of the path.
This non-uniqueness allows $\rho$ to also encode the {\it relative} orientation of such paths, i.e., if one path enclosing a singularity has associated parameter $\rho$, then a path with the opposite orientation enclosing the same singularity is naturally assigned the parameter $1/\rho$.
Thus the single complex parameter $\rho$ captures perfectly the monodromy of an isolated singularity.

\subsection{Explicit example of non-trivial monodromy}

An explicit example will now be used to demonstrate that it is quite normal for isolated singularities to have non-trivial monodromy, i.e. that isolated singularities will generally lead to the B\"acklund system (\ref{bt}) being inconsistent globally despite its local consistency.
Furthermore, examining this example will also suggest an alternative and perhaps more intuitive characterisation of this situation.

\begin{figure}[t]
\centering
\begin{picture}(260,120)(0,0)
% Long path picture
\multiput(10,2)(16,0){7}{{\nt\line(0,1){112}}}
\multiput(2,10)(0,16){7}{{\nt\line(1,0){112}}}
\multiput(58,42)(16,16){2}{\dt}
\multiput(42,58)(16,16){2}{\dt}
\put(58,58){\dt}
\put(42,58){\st\line(1,0){32}}
\put(58,42){\st\line(0,1){32}}
\multiput(58,10)(16,0){2}{\al}
\multiput(42,106)(16,0){2}{\ar}
\multiput(26,26)(0,16){4}{\au}
\multiput(90,42)(0,16){4}{\ad}
\multiput(42,26)(48,0){2}{\al}
\multiput(26,90)(48,0){2}{\ar}
\multiput(42,10)(0,80){2}{\au}
\multiput(74,26)(0,80){2}{\ad}
% Short path picture
\multiput(150,2)(16,0){7}{{\nt\line(0,1){112}}}
\multiput(142,10)(0,16){7}{{\nt\line(1,0){112}}}
\multiput(198,42)(16,16){2}{\dt}
\multiput(182,58)(16,16){2}{\dt}
\put(198,58){\dt}
\put(182,58){\st\line(1,0){32}}
\put(198,42){\st\line(0,1){32}}
\multiput(182,42)(0,16){2}{\au}
\multiput(182,74)(16,0){2}{\ar}
\multiput(214,74)(0,-16){2}{\ad}
\multiput(214,42)(-16,0){2}{\al}
\end{picture}
\caption{The simplest admissible singularity conf\/iguration. The f\/irst diagram shows a non-singular path enclosing the singularity.
The shortest such path is indicated in the second diagram. The paths can be deformed into each other by making alterations on a single quadrilateral at a time whilst avoiding singular edges.}
\label{acute}
\end{figure}
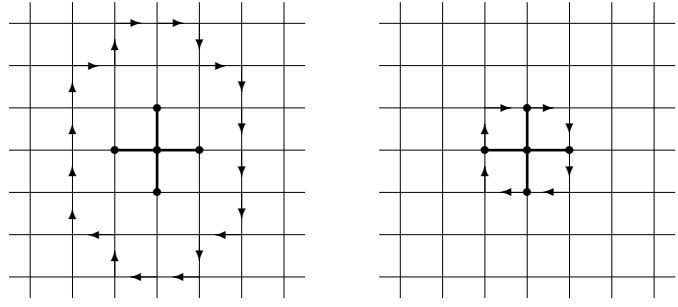

To exhibit non-trivial monodromy it is suf\/f\/icient to consider the simplest isolated singularity, which is the conf\/iguration of four edges in the form of a cross illustrated in Fig.~\ref{acute}.
Two non-singular paths enclosing the singularity are also shown.
Clearly it is convenient to consider the {\it shortest} enclosing path shown on the right; to calculate explicitly the monodromy using this path is a straightforward but quite lengthy calculation.
If the values of the function~$u$ on the nine points of the singular region are represented by the following matrix
\begin{gather*}
\left[ \begin{array}{ccc}
\wh{\wh{u}} & \wh{\wh{\wt{u}}} & \wh{\wh{\wt{\wt{u}}}} \\
\wh{u} & \wh{\wt{u}} & \wh{\wt{\wt{u}}} \\
u & \wt{u} & \wt{\wt{u}}
\end{array} \right],
\end{gather*}
then a solution of (\ref{ge}) with the indicated singularity in this region is represented by
\begin{gather}
\left[ \begin{array}{ccc}
c & f(\zeta_0+\beta) & d \\
f(\zeta_0+\alpha) & f(\zeta_0) & f(\zeta_0+\alpha) \\
a & f(\zeta_0+\beta) & b
\end{array} \right].
\label{cs}
\end{gather}
Here the function $f$ depends on which type-Q polynomial is being considered, this function was given in the right-hand column of Table \ref{Qpolys}.
It can be verif\/ied directly that (\ref{cs}) satisf\/ies (\ref{ge}) on each of the four quadrilaterals independently of the data
\begin{gather*}
\{a,b,c,d,\zeta_0\} %\label{sd}
\end{gather*}
which is therefore arbitrary.
Composing the eight transformations around the enclosing path results in a monodromy transformation which is the identity if and only if the following condition is satisf\/ied by this data:
\begin{gather}
[a+d-b-c][ad-f(\zeta_0+\alpha-\beta)f(\zeta_0-\alpha+\beta)] \nonumber\\
\qquad {} =    [ad-bc][a+d-f(\zeta_0+\alpha-\beta)-f(\zeta_0-\alpha+\beta)].
\label{toda}
\end{gather}
Therefore for generic data the singularity in Fig.~\ref{acute} has non-trivial monodromy.
Notice that condition~(\ref{toda}) is independent of the B\"acklund parameter~$\gamma$, in fact the singularity in Fig.~\ref{acute} belongs to an exceptional sub-class of conf\/igurations whose monodromy turns out to be independent of this parameter.

The condition (\ref{toda}) itself is quite interesting, in fact this condition is nothing but the discrete Toda system~\cite{abs1,as} which in the absence of singularities is a consequence of the equations on the quadrilaterals.
Therefore in this example there is a direct connection between asking if the solution satisf\/ies the larger-stencil formal consequences of~(\ref{ge}) and whether there is a singularity with non-trivial monodromy.

\subsection{Formal consequences of the equation}
To better understand the connection observed in the previous section it is useful to focus on the path which forms the external boundary of the singular region.
It is a corollary of Lemma \ref{absl3} that the following holds:
\begin{lemma}
Edges which lie on the boundary of the singular region are themselves non-singular.
\end{lemma}

 For instance the path indicated on the right of Fig.~\ref{acute} traverses the boundary of the singular region.
Clearly the monodromy of an isolated singularity can be computed by composing transformations along its external boundary.
Therefore the monodromy {\it must only depend on the values of the solution on this boundary}.
This leads to the following:
\begin{proposition}\label{eqv1}
Consider an isolated singularity in a solution of \eqref{ge}.
The constraints on the values of the solution on the external boundary of the singular region which arise when \eqref{ge} is used to eliminate variables on vertices interior to this boundary are sufficient for the monodromy to be trivial.
\end{proposition}

 The suf\/f\/iciency becomes clear by considering the same path, but with the singularity removed: in the absence of the singularity the formal constraints arising on the external boundary must be satisf\/ied, and the monodromy is of course trivial.
The trivial monodromy must therefore be a consequence of these constraints as the monodromy itself depends only on the values of the solution on the external boundary.

It would be natural to conjecture that the suf\/f\/iciency described in Proposition~\ref{eqv1} can be strengthened to suf\/f\/iciency {\it and} necessity.
It would require a more precise formulation, in particular demanding that the monodromy vanishes for {\it all values of the spectral parameter}.
Such a~proposition can in fact be proven for simple conf\/igurations, however counter-examples do exist in the case when the underlying system of polynomials are more degenerate than the \mbox{type-Q} systems considered here, thus any proof will rely on further analytic considerations of the polynomials and goes beyond the scope of the present paper.

Suf\/f\/ice it to say that imposing not just the quadrilateral equation~(\ref{ge}), but also the larger-stencil formal consequences of the equation, restricts solutions to those whose singularities have trivial monodromy.

\subsection{Smooth deformation of a non-singular solution}

There is a second point of view which in addition to Proposition~\ref{eqv1} provides a dif\/ferent but similarly intuitive way to understand of the dif\/ference between singularities with trivial and non-trivial monodromy.
It can be stated thus:
\begin{proposition}\label{eqv2}
If a solution of \eqref{ge} has been obtained from a non-singular solution by a~smooth limiting procedure, then any singularities present must have trivial monodromy.
\end{proposition}

The proof of this statement is very simple.
The monodromy around any path is of course trivial for the non-singular solution.
This remains true as the solution is deformed to approach the solution containing singularities.
In the limit the monodromy around any non-singular path must remain trivial because it is a smooth function of the values of the solution which are themselves being smoothly deformed.

Again, in the consideration of isolated singularities, it is natural to conjecture the converse of Proposition~\ref{eqv2}, but obtaining a proof is equal in complexity to proving the converse of Proposition~\ref{eqv1} and is beyond the scope of the present paper.

The Propositions \ref{eqv1} and \ref{eqv2} give alternative (but clearly connected) ways to understand the dif\/ference between isolated singularities which have trivial and non-trivial monodromy.
The reason these concepts are important is that, unlike the monodromy, they can be understood without f\/irst knowing about the integrability.

\subsection{The B\"acklund transformation and non-trivial lattice combinatorics}

\begin{figure}[t]
\centering
\begin{picture}(260,104)(0,0)
% necessarily acute singularity
\multiput(10,2)(16,0){7}{{\nt\line(0,1){96}}}
\multiput(2,10)(0,16){6}{{\nt\line(1,0){112}}}
\put(42,74){\dt}
\multiput(26,58)(16,0){4}{\dt}
\multiput(42,42)(16,0){4}{\dt}
\put(74,26){\dt}
\put(26,58){\st\line(1,0){32}}
\put(58,42){\st\line(1,0){32}}
\put(42,42){\st\line(0,1){32}}
\put(58,42){\st\line(0,1){16}}
\put(74,26){\st\line(0,1){32}}
% necessarily acute singularity and a path along its boundary
\multiput(150,2)(16,0){7}{{\nt\line(0,1){96}}}
\multiput(142,10)(0,16){6}{{\nt\line(1,0){112}}}
\put(182,74){\dt}
\multiput(166,58)(16,0){4}{\dt}
\multiput(182,42)(16,0){4}{\dt}
\put(214,26){\dt}
\put(166,58){\st\line(1,0){32}}
\put(198,42){\st\line(1,0){32}}
\put(182,42){\st\line(0,1){32}}
\put(198,42){\st\line(0,1){16}}
\put(214,26){\st\line(0,1){32}}
\multiput(166,42)(0,16){2}{\au}
\multiput(230,42)(0,16){2}{\ad}
\multiput(182,42)(16,0){2}{\al}
\multiput(166,74)(16,0){2}{\ar}
\multiput(214,26)(16,0){2}{\al}
\multiput(198,58)(16,0){2}{\ar}
\put(198,26){\au}
\put(198,74){\ad}
\end{picture}
\caption{An admissible singularity conf\/iguration which has non-trivial monodromy by virtue of its shape alone. The second diagram is the same as the f\/irst but includes a path traversing the boundary of the singular region.}
\label{veryacute}
\end{figure}
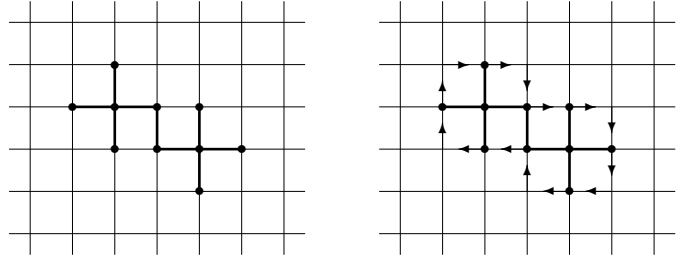

To apply the B\"acklund transformation it is quite natural to restrict attention to solutions whose singularities have trivial monodromy because the transformed solution is then well def\/ined on~$\mathbb{Z}^2$.
In this respect it is worth remarking that demanding an isolated singularity have trivial monodromy (for all values of the B\"acklund parameter) imposes conditions on its shape in addition to those of Section~\ref{SC}.
In other words there exist admissible singularity conf\/igurations which {\it necessarily} have non-trivial monodromy (for a generic value of the B\"acklund parameter).
Such a conf\/iguration is illustrated in Fig.~\ref{veryacute}.
What happens in this example is that imposing the trivial-monodromy conditions on the boundary forces some of the boundary edges to become singular, making the singular region change shape.

\begin{figure}[t]
\centering
\begin{picture}(240,240)(-120,-120)
\put(24.142,0){\line(1,0){90.532}}
\put(0,24.142){\line(0,1){90.532}}
\put(-24.142,0){\line(-1,0){90.532}}
\put(0,-24.142){\line(0,-1){90.532}}
\put(17.071,17.071){\line(1,1){64.016}}
\put(-17.071,17.071){\line(-1,1){64.016}}
\put(-17.071,-17.071){\line(-1,-1){64.016}}
\put(17.071,-17.071){\line(1,-1){64.016}}
\put(-10,-24.142){\line(1,0){20}}
\multiput(-10,-24.142)(20,0){2}{\line(0,-1){90.532}}
\put(-10,24.142){\line(1,0){20}}
\multiput(-10,24.142)(20,0){2}{\line(0,1){90.532}}
\put(-24.142,-10){\line(0,1){20}}
\multiput(-24.142,-10)(0,20){2}{\line(-1,0){90.532}}
\put(24.142,-10){\line(0,1){20}}
\multiput(24.142,-10)(0,20){2}{\line(1,0){90.532}}
\put(-24.142,10){\line(1,1){14.142}}
\multiput(-24.142,10)(14.142,14.142){2}{\line(-1,1){64.016}}
\put(10,-24.142){\line(1,1){14.142}}
\multiput(10,-24.142)(14.142,14.142){2}{\line(1,-1){64.016}}
\put(-10,-24.142){\line(-1,1){14.142}}
\multiput(-10,-24.142)(-14.142,14.142){2}{\line(-1,-1){64.016}}
\put(24.142,10){\line(-1,1){14.142}}
\multiput(24.142,10)(-14.142,14.142){2}{\line(1,1){64.016}}
\put(-20,-48.284){\line(1,0){40}}
\multiput(-20,-48.284)(40,0){2}{\line(0,-1){66.39}}
\put(-20,48.284){\line(1,0){40}}
\multiput(-20,48.284)(40,0){2}{\line(0,1){66.39}}
\put(-48.284,-20){\line(0,1){40}}
\multiput(-48.284,-20)(0,40){2}{\line(-1,0){66.39}}
\put(48.284,-20){\line(0,1){40}}
\multiput(48.284,-20)(0,40){2}{\line(1,0){66.39}}
\put(-48.284,20){\line(1,1){28.284}}
\multiput(-48.284,20)(28.284,28.284){2}{\line(-1,1){46.945}}
\put(20,-48.284){\line(1,1){28.284}}
\multiput(20,-48.284)(28.284,28.284){2}{\line(1,-1){46.945}}
\put(-20,-48.284){\line(-1,1){28.284}}
\multiput(-20,-48.284)(-28.284,28.284){2}{\line(-1,-1){46.945}}
\put(48.284,20){\line(-1,1){28.284}}
\multiput(48.284,20)(-28.284,28.284){2}{\line(1,1){46.945}}
\put(-30,-72.426){\line(1,0){60}}
\multiput(-30,-72.426)(60,0){2}{\line(0,-1){42.248}}
\put(-30,72.426){\line(1,0){60}}
\multiput(-30,72.426)(60,0){2}{\line(0,1){42.248}}
\put(-72.426,-30){\line(0,1){60}}
\multiput(-72.426,-30)(0,60){2}{\line(-1,0){42.248}}
\put(72.426,-30){\line(0,1){60}}
\multiput(72.426,-30)(0,60){2}{\line(1,0){42.248}}
\put(-72.426,30){\line(1,1){42.426}}
\multiput(-72.426,30)(42.426,42.426){2}{\line(-1,1){29.874}}
\put(30,-72.426){\line(1,1){42.426}}
\multiput(30,-72.426)(42.426,42.426){2}{\line(1,-1){29.874}}
\put(-30,-72.426){\line(-1,1){42.426}}
\multiput(-30,-72.426)(-42.426,42.426){2}{\line(-1,-1){29.874}}
\put(72.426,30){\line(-1,1){42.426}}
\multiput(72.426,30)(-42.426,42.426){2}{\line(1,1){29.874}}
\put(-40,-96.568){\line(1,0){80}}
\multiput(-40,-96.568)(80,0){2}{\line(0,-1){18.106}}
\put(-40,96.568){\line(1,0){80}}
\multiput(-40,96.568)(80,0){2}{\line(0,1){18.106}}
\put(-96.568,-40){\line(0,1){80}}
\multiput(-96.568,-40)(0,80){2}{\line(-1,0){18.106}}
\put(96.568,-40){\line(0,1){80}}
\multiput(96.568,-40)(0,80){2}{\line(1,0){18.106}}
\put(-96.568,40){\line(1,1){56.568}}
\multiput(-96.568,40)(56.568,56.568){2}{\line(-1,1){12.803}}
\put(40,-96.568){\line(1,1){56.568}}
\multiput(40,-96.568)(56.568,56.568){2}{\line(1,-1){12.803}}
\put(-40,-96.568){\line(-1,1){56.568}}
\multiput(-40,-96.568)(-56.568,56.568){2}{\line(-1,-1){12.803}}
\put(96.568,40){\line(-1,1){56.568}}
\multiput(96.568,40)(-56.568,56.568){2}{\line(1,1){12.803}}
\end{picture}
\caption{A quadrilateral lattice with combinatorics distinct from~$\mathbb{Z}^2$. It is the lattice on which the transformed solution is def\/ined if the B\"acklund transformation is applied to a solution with the singularity conf\/iguration of Fig.~\ref{acute} in the case that the square of the monodromy transformation is the identity.}
\label{octo}
\end{figure}
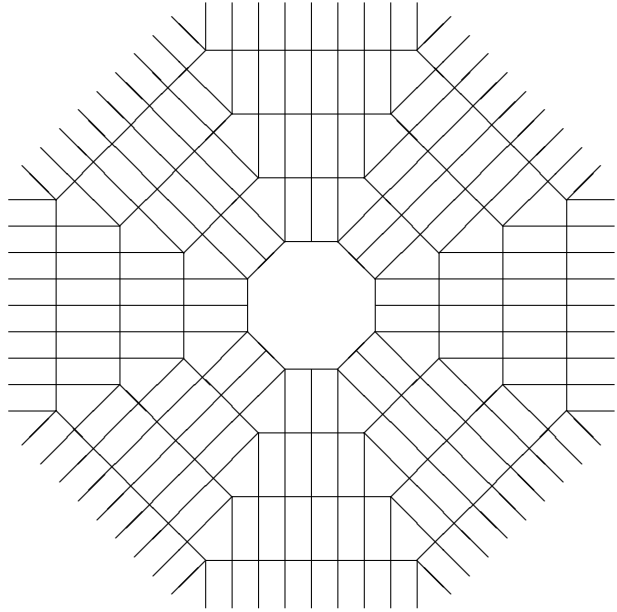

However it is not necessary to restrict attention only to solutions whose singularities have trivial monodromy.
Whilst the transformed solution may not be well def\/ined on the $\mathbb{Z}^2$ lattice, it {\it is} well def\/ined on a lattice with dif\/ferent combinatorics.
An example of this can be exhibited by returning to the conf\/iguration shown in Fig.~\ref{acute}.
We suppose the monodromy is non-trivial, but will consider the simplest non-trivial situation which occurs when the square of the monodromy transformation is the identity, in particular this means that $\rho=-1$.
The transformed solution is not def\/ined on $\mathbb{Z}^2$, the actual lattice on which it is uniquely def\/ined is illustrated in Fig.~\ref{octo}.
This lattice can be viewed as two copies of $\mathbb{Z}^2$ with the vertex at the centre of the singularity removed, and which are then joined together along a {\it branch cut} along a path connecting the removed vertex with a point at inf\/inity.
The lattices should be joined in such a way that a path once around the removed vertex causes a switch from one lattice onto the other.
The new lattice could be visualised as the embedding of a two-dimensional lattice in three dimensions, Fig.~\ref{octo} gives an alternative way to view the lattice~-- the combinatorics are dif\/ferent but it is embedded in just two dimensions.

Thus it is natural that in the presence of singularities the B\"acklund transformation connects solutions of the same equation, but on domains with dif\/ferent underlying combinatorics.
The precise combinatorics are a function of the singularity conf\/iguration and the monodromy.
The procedure of def\/ining the domain of the transformed solution is analogous to analytic conti\-nua\-tion.

In conclusion, the restriction to solutions whose singularities have trivial monodromy means the B\"acklund transformation can be considered as usual: one iterate into a new direction on the regular multidimensional lattice.
Lifting this restriction leads to an interesting situation that can also be described on a higher dimensional lattice, but with the dif\/ference that the combinatorics of the lattice will depend on the position, shape and monodromy of the singularities.

\subsection{Remark on the monodromy in multidimensions}

The important generalisation to multidimensions requires a fresh look at the notion of mo\-no\-dromy.
We start with the B\"acklund transformation in multidimensions; given a solution $u$ of the $N$-dimensional system (\ref{mde}) the B\"acklund equations for a new solution $v$ are
\begin{gather}
\cB_i(u,\oT_iu,v,\oT_iv)=0, \qquad i\in\{1,\ldots, N\}\label{gbt},
\end{gather}
where $\cB_i=\cP|_{(\alpha,\beta)\rightarrow(\alpha_i,\gamma)}$.
The relationship between the B\"acklund transformation and the singular edges is a direct extension of the two-dimensional situation:
the equations (\ref{gbt}) associate M\"obius transformations $v\mapsto\oT_iv$ to all edges of the multidimensional lattice {\it except those on which solution $u$ is singular}.

To see how the idea of monodromy extends to situations beyond $N=2$ suppose f\/irst that \mbox{$N=3$}.
It is clear that isolated singularities in three dimensions do not lead to global inconsistency of~(\ref{gbt}) because they make no obstruction to gradual deformation of one lattice path into another.
The singularities which cause obstruction to deformations of this kind in three dimensions include singularities which form a {\it closed loop}, or which are {\it infinite in extent in one dimension}.
That such singularities could have non-trivial monodromy relies on the fact pointed out in Section~\ref{MON}: the trivial monodromy at a f\/ixed value of the spectral parameter which is necessary for the extension of a two-dimensional solution to three dimensions, does not mean the monodromy is trivial for all values of the parameter.

In general then, the class of singularities with the potential to have non-trivial monodromy needs to be considered in the context of the {\it dimension} of the underlying system.
Only in two dimensions are the singularities with non-trivial monodromy necessarily {\it isolated}.

\subsection*{Acknowledgements}
The author gratefully acknowledges helpful discussions with Nalini Joshi.
The research was funded by Australian Research Council Discovery Grants DP 0985615 and DP 110104151.

\pdfbookmark[1]{References}{ref}
\LastPageEnding

\begin{thebibliography}{99}

\footnotesize\itemsep=0pt

\bibitem{abs1}
Adler V.E., Bobenko A.I., Suris Yu.B.,
Classif\/ication of integrable equations on quad-graphs. The consistency approach,
\href{http://dx.doi.org/10.1007/s00220-002-0762-8}{{\it Comm. Math. Phys.}} {\bf 233} (2003), 513--543,
\href{http://arxiv.org/abs/nlin.SI/0202024}{nlin.SI/0202024}.

\bibitem{nw}
Nijhof\/f F.W., Walker A.J.,
The discrete and continuous Painlev\'e VI hierarchy and the Garnier systems,
\href{http://dx.doi.org/10.1017/S0017089501000106}{{\it Glasg. Math.~J.}} {\bf 43A} (2001), 109--123,
\href{http://arxiv.org/abs/nlin.SI/0001054}{nlin.SI/0001054}.

\bibitem{bs}
Bobenko A.I., Suris Yu.B.,
Integrable systems on quad-graphs,
\href{http://dx.doi.org/10.1155/S1073792802110075}{{\it Int. Math. Res. Not.}} {\bf 2002} (2002), no.~11, 573--611,
\href{http://arxiv.org/abs/nlin.SI/0110004}{nlin.SI/0110004}.

\bibitem{ahn}
Atkinson J., Hietarinta J., Nijhof\/f F.W.,
 Seed and soliton solutions for Adler's lattice equation,
\href{http://dx.doi.org/10.1088/1751-8113/40/1/F01}{{\it J.~Phys.~A: Math. Theor.}} {\bf 40} (2007), F1--F8,
\href{http://arxiv.org/abs/nlin.SI/0609044}{nlin.SI/0609044}.

\bibitem{abs2}
Adler V.E., Bobenko A.I., Suris Yu.B.,
Discrete nonlinear hyperbolic equations: classif\/ication of integrable cases,
\href{http://dx.doi.org/10.1007/s10688-009-0002-5}{{\it Funct. Anal. Appl.}} {\bf 43} (2009), 3--17,
\href{http://arxiv.org/abs/0705.1663}{arXiv:0705.1663}.

\bibitem{adl}
Adler V.E.,
B\"acklund transformation for the Krichever--Novikov equation,
\href{http://dx.doi.org/10.1155/S1073792898000014}{{\it Int. Math. Res. Not.}} {\bf 1998} (1998), no.~1, 1--4,
\href{http://arxiv.org/abs/solv-int/9707015}{solv-int/9707015}.

\bibitem{av}
Adler V.E., Veselov A.P.,
Cauchy problem for integrable discrete equations on quad-graphs,
\href{http://dx.doi.org/10.1007/s10440-004-5557-9}{{\it Acta Appl. Math.}} {\bf 84} (2004), 237--262,
\href{http://arxiv.org/abs/math-ph/0211054}{math-ph/0211054}.

\bibitem{dsss}
Dolbilin N.P., Sedrakyan A.G., Shtan'ko M.A., Shtogrin M.I.,
Topology of a family of parametrizations of two-dimensional cycles arising in the three-dimensional Ising model,
{\it Dokl. Akad. Nauk SSSR} {\bf 295} (1987), no.~1, 19--23 (English transl.: {\it Soviet Math. Dokl.} {\bf 36} (1988), no.~1, 11--15).

\bibitem{boll}
Boll R.,
Classif\/ication of 3D consistent quad-equations,
\href{http://arxiv.org/abs/1009.4007}{arXiv:1009.4007}.

\bibitem{as}
Adler V.E., Suris Yu.B.,
${\rm Q}_4$: integrable master equation related to an elliptic curve,
\href{http://dx.doi.org/10.1155/S107379280413273X}{{\it Int. Math. Res. Not.}} {\bf 2004} (2004), no.~47, 2523--2553,
\href{http://arxiv.org/abs/nlin.SI/0309030}{nlin.SI/0309030}.

\end{thebibliography}
\end{document}